# Gate-controlled magnetic phase transition in a van der Waals magnet Fe$_5$GeTe$_2$

Cheng Tan[1#], Wen-Qiang Xie[2#], Guolin Zheng[1*], Nuriyah Aloufi[1], Sultan Albarakati[1], Meri Algarni[1], Jiangpeng Song[3], James Partridge[1], Dimitrie Culcer[4], Xiaolin Wang[5,6], Jiabao Yi[7], Yimin Xiong[3], Mingliang Tian[3,8,9], Yu-Jun Zhao[2*], Lan Wang[1*]

[1]School of Science, RMIT University, Melbourne, VIC 3001, Australia.

[2]Department of Physics, South China University of Technology, Guangzhou 510640, China.

[3]Anhui Province Key Laboratory of Condensed Matter Physics at Extreme Conditions, High Magnetic Field Laboratory, Chinese Academy of Sciences (CAS), Hefei 230031, Anhui, China.

[4]School of Physics and ARC Centre of Excellence in Future Low-Energy Electronics Technologies, UNSW Node, University of New South Wales, Sydney, NSW 2052, Australia.

[5]Institute for Superconducting & Electronic Materials, Australian Institute of Innovative Materials, University of Wollongong, Wollongong, NSW 2500, Australia.

[6]ARC Centre for Future Low-Energy Electronics Technologies (FLEET), University of Wollongong, Wollongong, NSW 2500, Australia.

[7]Global Innovative Center for Advanced Nanomaterials, School of Engineering, University of Newcastle, Callaghan NSW 2308, Australia.

[8]Department of Physics, School of Physics and Materials Science, Anhui University, Hefei 230601, Anhui, China.

[9]Collaborative Innovation Center of Advanced Microstructures, Nanjing University, Nanjing 210093, China.

[#] Those authors equally contribute to the paper.

[*] Corresponding authors. Correspondence and requests for materials should be addressed to G. Zheng (email: guolin.zheng@rmit.edu.au), Y.-J. Z. (email: zhaoyj@scut.edu.cn) and L. W. (email: lan.wang@rmit.edu.au).

# Abstract

Magnetic van der Waals (vdW) materials, including ferromagnets (FM) and antiferromagnets (AFM), have given access to the investigation of magnetism in two-dimensional (2D) limit and attracted broad interests recently. However, most of them are semiconducting or insulating and the vdW itinerant magnets, especially vdW itinerant AFM, are very rare. Here, we studied the anomalous Hall effect of a vdW itinerant magnet $Fe_5GeTe_2$ (F5GT) with various thicknesses down to 6.8 nm (two unit cells). Despite the robust ferromagnetic ground state in thin-layer F5GT, however, we show that the electron doping implemented by a protonic gate can eventually induce a magnetic phase transition from FM to AFM. Realization of an antiferromagnetic phase in F5GT highlights its promising applications in high-temperature antiferromagnetic vdW devices and heterostructures.

# MAIN TEXT

**Introduction**

The emergence of two-dimensional (2D) magnetic materials (*1-3*) has expedited the novel vdW spintronic devices such as tunneling magnetoresistance (*4-7*), giant magnetoresistance (*8*) and spin-orbit torque (*9, 10*). Compared to itinerant ferromagnets (FMs), antiferromagnets (AFMs) have their unique advantages as building blocks of the spintronic devices. For example, AFM is robust to the stray magnetic fields, making them suitable for memory devices (*11, 12*); the required current density in AFM-based spin-orbit torque devices is much lower than that in FM, indicating their much lower energy consumption (*13-17*). Regardless of the recent rise of vdW magnets, vdW itinerant antiferromagnets are very scarce (*18, 19*), most of the emerged vdW antiferromagnets are either insulating or semiconducting (*2, 20-24*). Therefore, the realization of vdW itinerant antiferromagnet with higher Néel temperature is still challenging.

Here, we systematically investigated the anomalous Hall effect in a vdW itinerant ferromagnet $Fe_5GeTe_2$ (F5GT) under different thicknesses and gating voltages and showed that magnetic ground state in F5GT is sensitive to the doping level. Despite its robust ferromagnetic ground state in thin-layer F5GT, electron doping can substantially suppress the FM and eventually induce a magnetic phase transition from FM to AFM in F5GT nanoflakes. Further theoretical calculation based on density functional theory (DFT) supports the experimental observations and demonstrates that the interlayer ferromagnetic coupling energy of F5GT can be substantially boosted by electron doping, resulting in an antiferromagnetic ground state. Controlling of the magnetic phase by gate voltages in F5GT not only opens up the opportunity of building antiferromagnetic spintronic devices at high temperatures, but also provides a new train of thought in searching for the vdW itinerant AFMs.

**Results**

F5GT is a newly synthesized vdW ferromagnet (*25-29*) which outstands for its higher $T_C$ at room temperatures and surpasses another widely investigated itinerant vdW FMs Fe$_3$GeTe$_2$ (FGT) (*3, 30-33*) with $T_C \approx 200\ K$. Fig. 1A illustrates a schematic diagram of the Fe$_5$GeTe$_2$ crystal structure projected along [100] direction. Similar to FGT, F5GT crystal is formed by thicker Fe-Ge slabs sandwiched by Te layers. The crystal structure of F5GT has a space group R$\bar{3}$m with lattice parameters a=4.04 Å and c=29.19 Å (*25*). In each unit cell, there are 3 Fe sites, and Fe$^1$ site is regarded as a split site located either above or below the Ge site, leading to the complex crystal structure. Fig.1B shows the optical and the atomic force microscope image of an ultrathin F5GT nanoflake on a SiO$_2$ substrate. Figure 1C illustrates the lateral distance dependent height curve (corresponding to the dashed blue line in Fig. 1B) of a thin F5GT nanoflake, indicating a thickness of 6.8 nm (2 u.c.). The normalized resistance temperature (RT) curves of F5GT devices with various thicknesses are shown in Fig. 1D. For thicker samples above 6 u.c., the RT curves exhibit a kink around $T \approx 145$ K. This resistance anomaly was attributed to the gradual magnetic transition of the Fe$^1$ site with decreasing the temperature (*25*). Additionally, a small response near 115 K is observed in the magnetization data collected from bulk crystal at 1 T (Supplementary Fig. S1a), which indicates the first order transition of the magneto-structure as well (*25, 26*). Moreover, the kink temperature is elevated up to 181 K and 203 K for 12 nm (4 u.c.) and 17 nm (6 u.c.) nanosheets, respectively. The dramatic elevation of the kink temperature in thinner nanosheets reveals that FM can be effectively modulated by thickness in F5GT which will be discussed later.

Figure 2A illustrates the magnetic field dependent anomalous Hall resistivity $\rho_{xy}$ of F5GT nanosheets with different thicknesses. The magnetic field is oriented perpendicular to the sheet plane for all samples. The 39 u.c. nanosheet displays several soft magnetic phases with a coercive field of 65.5 mT, which agrees well with the magnetic measurement of the single crystal (Supplementary Fig. S1). Reducing the thickness down to 14 u.c., the anomalous Hall resistivity ratios $\rho_{xy}(0\ T)/\rho_{xy}^{SAT}$ (with $\rho_{xy}^{SAT}$ the saturated anomalous Hall resistivity) surges from 0.24 in

39 u.c. nanosheet to 0.73 in 14 u.c. nanosheet and correspondingly, the coercive field elevates up to 93.8 mT in 14 u.c. nanosheet. However, the magnetic loops for nanosheets below 14 u.c. are nearly square-shaped, indicating a strong perpendicular magnetic anisotropy, as can be seen in Fig. 2A. Ramping the magnetic field from 1 T to -1 T, the hysteresis loop of 6 u.c. nanosheet exhibits a nearly square-shaped loop with a slender tail near the coercive field $H_c$. This tail in magnetic loop near $H_c$ unveils the co-existence of two magnetic phases in 6 u.c. nanosheet. While in 12 nm (4 u.c.) nanosheet, it shows two hard magnetic phases with different coercivities. Further lowering the thickness down to 6.8 nm (2 u.c.), it finally shows a single, hard magnetic phase with a dramatically large $H_c$ of 1.4 T and a $\rho_{xy}(0\,T)/\rho_{xy}^{SAT}$ ratio of 1.02. The larger $\rho_{xy}(0\,T)/\rho_{xy}^{SAT}$ ratios indicate the magnetic moments align perpendicular to the nanosheet plane at the remanence point in thinner F5GT nanosheets. Generally, magnetic anisotropy is mainly determined by the magnetocrystalline anisotropy and shape anisotropy, and should not change too much as the decreasing of the thickness. The increasing $\rho_{xy}(0\,T)/\rho_{xy}^{SAT}$ values should be attributed to the weaker interlayer magnetic domain interactions with decreasing the thickness. Compared to FGT nanosheets of the similar thicknesses (31), thicker F5GT nanosheets (>3 u.c.) exibit much smaller coercivities. The different coercivities in F5GT and FGT should be ascribed to the different interlayer magnetic coupling and magnetic domain wall (the boundary-region between two magnetic domains with different magnetization orientations) motion around the coercive fields. When the field is approaching the coercive field, a flipped magnetic domain can affect the domains in both inter-and intra-layers due to the magnetic coupling, leading to a small coercivity. Since the interlayer magnetic coupling in F5GT is stronger than that in FGT, it is understandable that thicker F5GT has relatively smaller coercivities than FGT of similar thicknesses due to the collective motion of coupled magnetic domain walls. Additionally, when the thickness of F5GT decreases further to 2 u.c., the lack of magnetic domain wall motion increases the threshold of the total magnetization inversion, leading to a much larger coercivity. Figure 2B presents the temperature-dependent remanence of F5GT nanosheets of different thicknesses. For nanosheets above 4 u.c., the remanence firstly increases with decreasing the temperature and reaches the maximum at around 120-150 K, followed by a sharp decline. This indicates a possible ferrimagnetic phase in thicker F5GT nanosheets rather than a ferromagnet (*25-27*). Note that, the remanence of 4 u.c. F5GT nanosheet decreases to zero at around 266.4 K (Supplementary Fig. S3), revealing a lower $T_C$ in 4 u.c. F5GT. The lower $T_C$ might be caused by weaker interlayer magnetic coupling due to the decreasing of the layer numbers, akin to some other vdW magnets (*1, 3, 31*). For 2 u.c. F5GT, however, the temperature-dependent remanence behaves as a typical ferromagnet below 220K, which is in accordance with the data in Fig.1D and Fig. 2A. As discussed above, the magnetic transition on the Fe[1] site results in resistance kinks on the temperature dependent $R_{xx}$ curves and the kink temperature varies with thicknesses. Fig. 2C highlights the thickness dependent $\rho_{xy}(0\ T)/\rho_{xy}^{SAT}$ and $T_k$, the definition of $T_k$ is illustrated in the Supplementary Fig. S5. It shows that the $\rho_{xy}(0\ T)/\rho_{xy}^{SAT}$ values increase with decreasing the thicknesses, indicating the magnetic anisotropy gradually aligns perpendicular to the sample plane in thinner nanoflakes. The $T_k$ values are typically around 150 K and almost unchanged for the nanosheets ranging from 40 nm (14 u.c.) to 120 nm (40 u.c.). However, this value abruptly jumps up to above 180 K when the thickness decreases down to 20 nm (7 u.c.).

Compared to FGT, the recent works reveal that the magnetic ground state in F5GT is sensitive to Cobalt substitution (*34-36*). However, it is unclear whether the magnetic ground state in F5GT is also sensitive to the *in-situ* charge doping. Previous experiments demonstrate that magnetism in vdW materials can be controlled by a gate voltage (*37-40*), which sheds lights on their applications in vdW spintronics and memory devices. For vdW itinerant magnet, however, conventional electric-field gating is stumbling in the electrical control of the FM, since the electric field tends to be screened within few nano-meters. Up to date, gate-induced Lithium ion doping (*3*) and protonic intercalation (*41*) have been proved to be effective ways to tune the magnetism and interlayer coupling in vdW itinerant magnet FGT. Using the same protonic gate technique, as illustrated in Fig. 3A, we find the FM in F5GT nanosheets can be starkly modulated. Fig. 3B shows the magnetic field dependent anomalous Hall resistivity $\rho_{xy}$ at various gate voltages for the 40 nm (14 u.c.) F5GT at 2 K. In pristine nanosheet, the hysteresis loop shows a large $\rho_{AHE}$ value, with $\rho_{AHE}$ defined by $\rho_{AHE} = |\rho_{xy}(1\,T) - \rho_{xy}(-1\,T)|$. Sweeping the voltage from 0 V to $V_g$= -3.1 V and -3.6 V, we find the $\rho_{AHE}$ decreases accordingly. The anomalous Hall resistivity can usually be written as $\rho_{xy} = R_0 H + \rho_{xy}^A = R_0 H + R_s M_z$, where $M_Z$ represents the magnetization (*42*). As the normal Hall section $R_0 H$ is significantly small compared with the whole anomalous Hall resistance, the hysteresis loop here is proportional to the magnetization loop. So the decrease of $\rho_{xy}$ indicates the decline of the magnetization. Note that the anomalous Hall exhibits a sign reversal at $V_g$= -4.2 V and -4.5 V during this process, the reversal of hysteresis loops at -4.2 V and -4.5 V indicate the reversal of $R_s M_Z$. It is possible that the sign reversal of $\rho_{xy}^A$ is merely ascribed to the flip of total magnetization $M_Z$ due to the change of carrier type. However, there is no obvious experimental evidence of sign change of Hall slope around -4.5 V, this sign reversal of $\rho_{xy}^A$ cannot be simply attributed to the change of carrier type. To procure more insights into the sign reversal of anomalous Hall near -4.5 V, we calculated the electron doping range-separated hybrid functional HSE06 (*43*) band structure of F5GT and its corresponding anomalous Hall conductivity in terms of Berry curvature (*42, 44*), as shown in Supplementary Fig. S7. The Hall conductivity ($\sigma_{xy}^z$) indeed exhibits a sign reverse in conduction band in Fig. S7d, indicating that an electron doping treatment of the system can lead to a Hall conductivity reversal. The calculated $\sigma_{xy}^z$ is around 5 $\Omega^{-1}\cdot cm^{-1}$, which is consistent with the experimental value obtained at $V_g$= -4.2 V and -4.5 V (5~10 $\Omega^{-1}\cdot cm^{-1}$). Note that the anomalous Hall curves are nearly 'flat' at high magnetic fields for nanosheets thinner than 14 u.c.. In order to determine the doping type, we tested more samples with different thicknesses (Supplementary Fig. S4). For 78 nm (27 u.c.) nanosheet, we found that negative gate voltages lead to an electron-type doping (Supplementary Fig. S4a) which correspondingly suppresses the anomalous Hall resistivity. A similar suppression of anomalous Hall resistivity has also been identified in other samples at negative gate voltages, indicating the same n-type doping at negative gate voltages. Increasing the gate voltage again, the hysteresis loops are further supressed and eventually disappears at $V_g$ = -5 V, implying a magnetic phase transition or magnetic anisotropy change. Since the magnetic anisotropy is mainly determined by magnetocrystalline anisotropy, it is unlikely that the magnetic anisotropy in F5GT has changed from out-of-plane to in-plane direction due to the proton intercalations. Additionally, F5GT shows the paramagnetic behaviors near room temperatures (>250 K, Supplementary Fig. S2), indicating a large energy gap (>20 meV) between the FM state and paramagnetic state at 2 K. Therefore, the disappeared hysteresis loop at $V_g$ = -5 V can probably be attributed to the formation of an AFM state. Fig. 3C shows the $V_g$ dependent electron carrier density of 27 u.c. nanosheet as well as the anomalous Hall resistivity ratio $\rho_{AHE}/\rho_{AHE}(0\,V)$ of 14 u.c. and 27 u.c. nanosheets, respectively. Sweeping the voltage $V_g$ from 0 to -5 V, $\rho_{AHE}/\rho_{AHE}(0\,V)$ in both 14 u.c. and 27 u.c. nanosheets decrease and electron carrier density in 27 u.c. nanosheet increases by about $0.65\times10^{22}$ cm$^{-3}$ from 0 V to -5 V. As discussed above, since $\rho_{AHE}/\rho_{AHE}(0\,V)$ of the 14 u.c. nanosheet exhibits a similar tendency under negative gate voltages, it is reasonable to infer that electron-doping suppresses ferromagnetism and probably gives rise to a magnetic phase transition from FM to AFM in F5GT nanosheets.

**Discussion**

To further judge and confirm the experimental results above, density functional theory (DFT) calculations were carried out (*45-48*). We found that the energy difference between FM coupling and AFM coupling is very small in F5GT, which may explain the dramatic evolution of hysteresis loop under various gate voltages. Fig. 4A shows the charge doping dependent energy difference between FM and AFM. Here the $\Delta E$ is defined as $\Delta E = E_{FM} - E_{AFM}$, where $E_{AFM}$ and $E_{FM}$ represents the energy of antiferromagnetic and ferromagnetic F5GT, respectively. In Fig. 4A, both PBE and LDA+U functionals exhibit a decreasing of $\Delta E$ with increasing hole doping (i.e decreasing electron doping), implying that the charge doping process can greatly affect the magnetic order in F5GT. This is in line with our experimental observation that the $\rho_{AHE}/\rho_{AHE}(0\,V)$ values consecutively decrease with the increasing of the gating voltages. Besides, the calculated ΔE of both LDA+U and PBE method is only -1.42 meV and 1.45 meV, respectively. Combined with the large energy gap (>20 meV) between the FM state and the paramagnetic state discussed above, the disappearance of anomalous Hall loop at $V_g$=-5 V in 14 u.c. F5GT nanoflake is predominately ascribed to the magnetic phase transition from FM to AFM in F5GT induced by electron doping. Fig. 4B compares the density of states (DOS) of FM/AFM F5GT under LDA+U and PBE functionals. Generally, charge doping has little effect on a material's total DOS. For F5GT, the total DOS consists of FM DOS and AFM DOS. As shown in Fig. 4A, when the F5GT is in FM state, *ΔE* (absolute value) drops prominently with the increasing of the electronic doping. Under this condition, part of FM coupling in F5GT transfers to AFM coupling, leading to a decreasing total DOS of the system.

In conclusion, we studied the anomalous Hall effect in a vdW itinerant ferromagnet F5GT under different thicknesses and gating voltages and found that magnetic ground state in F5GT can be *in-situly* tuned by a gate voltage. Electron doping in F5GT induced by a protonic gate can trigger a magnetic phase transition from FM to AFM. Theoretical calculation based on density functional theory supports the experimental observations and demonstrates that the interlayer ferromagnetic coupling energy of F5GT can be enhanced by electron doping, leading to an antiferromagnetic ground state. Realizing of AFM phase in near room temperature vdW ferromagnet F5GT nanosheets by protonic gating enables probable vdW antiferromagnetic devices and heterostructures at high temperatures.

## Materials and Methods

### Single-crystal growth

Part of the single crystals were bought from HQ graphene, while others were grown by transport method as discussed in ref. 25.

### Device fabrication

F5GT flakes of various thicknesses were mechanically exfoliated onto $SiO_2$/Si wafers in a glove box with oxygen and water levels below 0.1 parts per million. The nanoflakes were examined by an optical microscope in glove box. Atomically smooth flakes were identified and then transferred via a Polycarbonates based pick-up method onto pre-patterned Pt electrodes or proton conductor for the following electron beam lithography.

### Electrical and magnetic measurement

The transport and magnetic measurements were performed in a MPMS3 SQUID (Superconducting Quantum Interference Device) magnetometer (Quantum Design, San Diego, CA, USA) with a base temperature of 1.8 K and a magnetic field of up to 7 T.

## Supplementary Materials

Fig. S1 Magnetic measurements of F5GT single crystal.

Fig. S2 Detailed hysteresis behaviors of 2 u.c. and 39 u.c. F5GT devices.

Fig. S3 Definition of the temperature dependent remanence curve for 2 u.c. device.

Fig. S4 Protonic gating modulation of other 3 devices at 2 K.

Fig. S5 Definition of the Fe1 site magnetic transition temperature Tk.

Fig. S6 Angle dependent anomalous Hall measurement of 2 u.c. and 5 u.c. devices.

Table S1. Density functional theory calculations with different lattice parameters and methods.

Fig. S7 HSE06 band structures and Hall conductivity with electron doping.

**Acknowledgments**

**General**: This research was performed in part at the RMIT Micro Nano Research Facility (MNRF) in the Victorian Node of the Australian National Fabrication Facility (ANFF) and the RMIT Microscopy and Microanalysis Facility (RMMF).

**Funding:** This research was supported by the Australian Research Council Centre of Excellence in Future Low-Energy Electronics Technologies (CE170100039), the Natural Science Foundation of China (11574088), the National Key Research and Development Program of China (2016YFA0300404).

**Competing interests:** The authors declare that they have no competing interests.

**Data and materials availability:** All data needed to evaluate the conclusions in the paper are present in the paper and/or the Supplementary Materials. Additional data related to this paper may be requested from the authors.

**Figures and Tables**

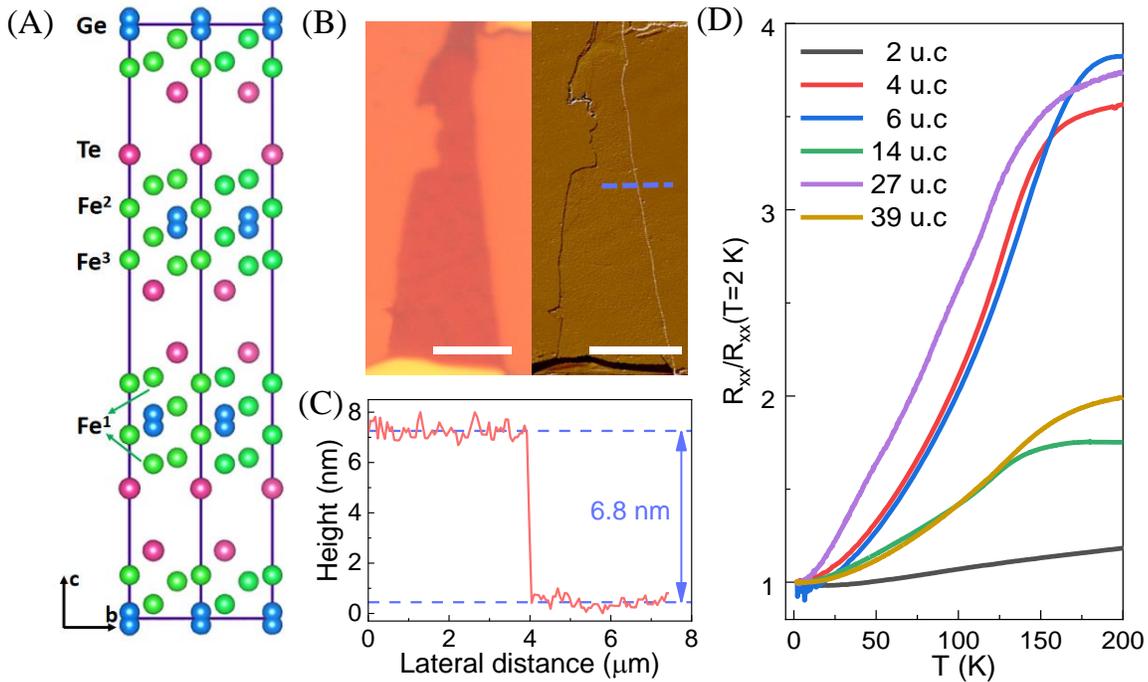

**Fig. 1 Crystal structure and initial characterization of Fe₅GeTe₂.** (**A**) Schematic diagram of the F5GT crystal structure viewed along a-axis. (**B**) Optical and atomic force microscope image of a thin F5GT nanoflake. The white scale bars represent 5 μm. (**C**) Cross-sectional profile of the F5GT flake along the dashed line in (B). (**D**) Temperature dependent longitudinal resistance of F5GT devices with a varying number of layers. Resistances are normalized to their values at $T=2$ K.

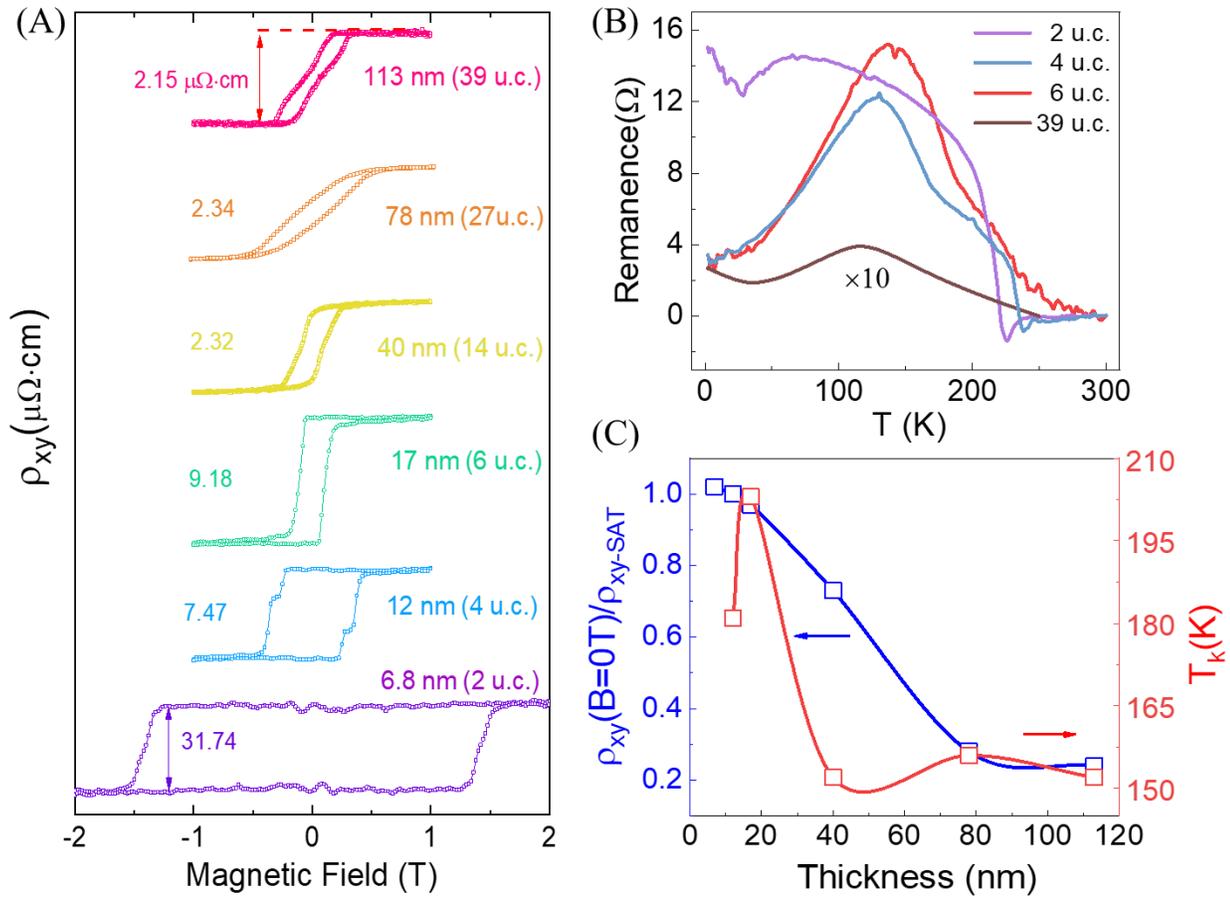

**Fig. 2 Anomalous Hall effect measurements performed on Fe₅GeTe₂ devices of various thicknesses.** (A) $\rho_{xy}(B)$ for F5GT nanoflakes of various thicknesses at 2 K. (B) Temperature dependent remanence curves of four F5GT devices. (C) Thickness dependent $\rho_{xy}(B=0\ T)/\rho_{xy}^{SAT}$ and $T_k$ curves.

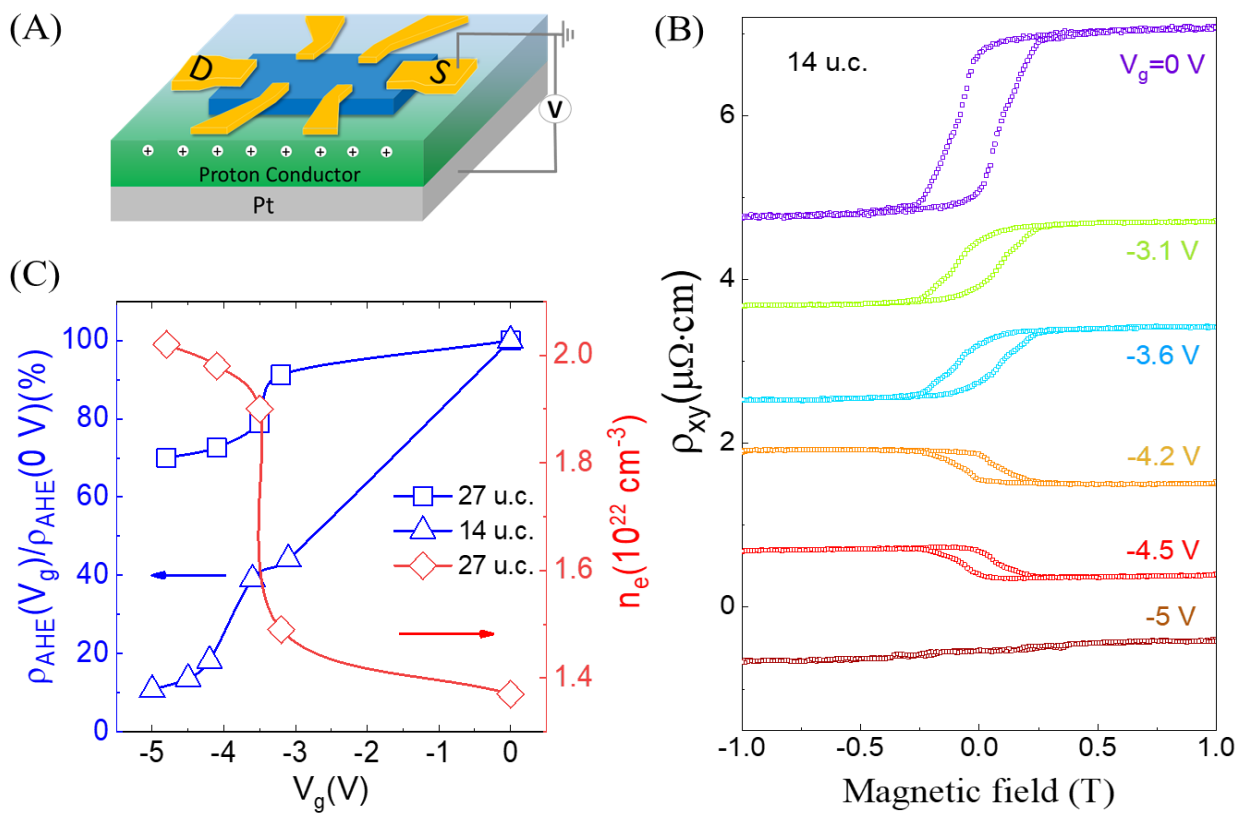

**Fig. 3 Protonic gating modulation of the 14 u.c. Fe₅GeTe₂ nanosheet.** (**A**) Schematic diagram of a protonic gating device. A Pt electrode is used to inject or pull protons out of the device. (**B**) $\rho_{xy}(B)$ loops at various gating voltages on the 14 u.c. F5GT device. (**C**) Gate voltage dependent carrier densities and anomalous Hall change ratio of the 14 u.c. and 27 u.c. F5GT nanosheets. The definition of the $\rho_{AHE}$ is the same as that in Fig. 2A.

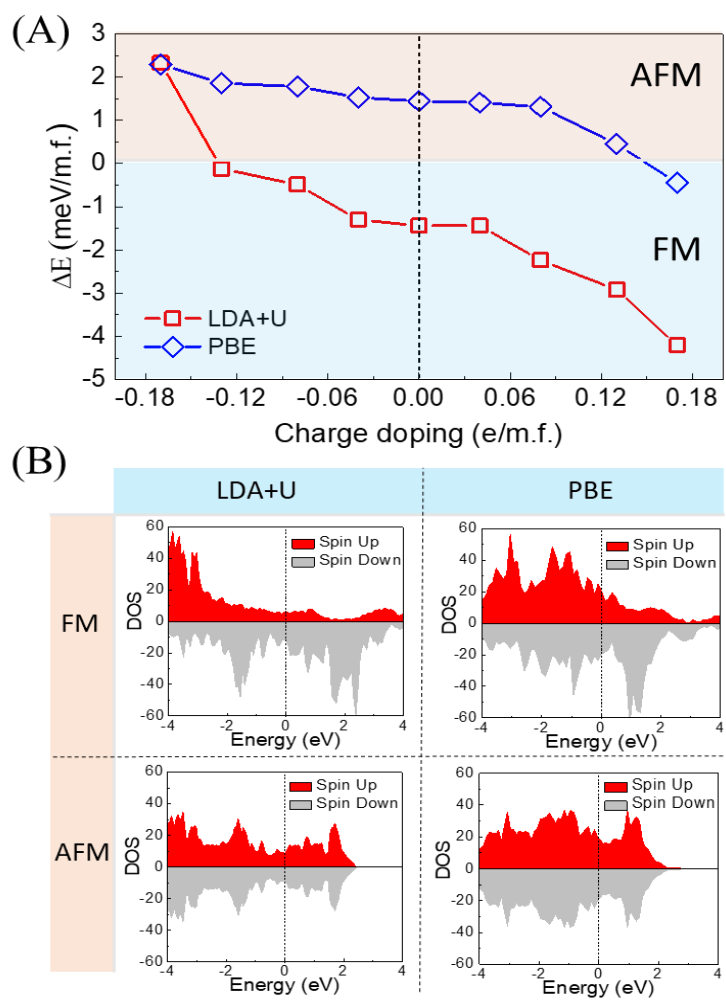

**Fig. 4 Theoretical calculation**. **(A)** Evolution of energy difference between FM and AFM (*ΔE*) with the charge doping under LDA+U and PBE functionals. The ochre region indicates AFM coupling, while the blue region implies FM coupling. **(B)** The total DOS of FM/AFM of F5GT under LDA+U and PBE functionals. The red region indicates up-spin DOS, while grey region implies down-spin DOS.

# Supplementary Materials

**Section S1. Magnetic measurements**

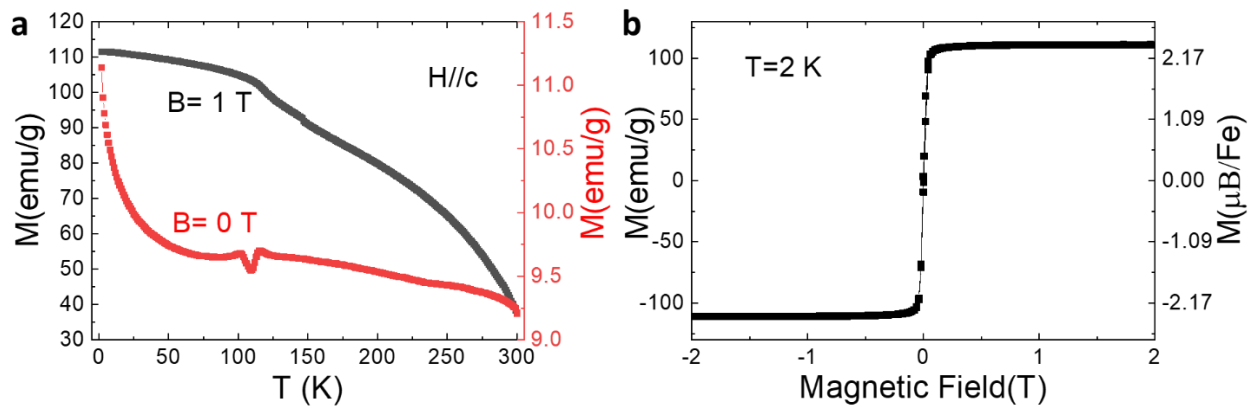

**Fig. S1 Magnetic measurements of F5GT single crystal.** **(a)** Temperture dependence of magnetization for F5GT under B= 1 T and B= 0 T. B= 0 T curve is measured by decreasing field from 1 T to 0 T at 2 K, then scanning up the temperature. **(b)** Magnetic field dependent magnetization measured at 2 K.

**Section S2. Detailed transport measurements for 2 u.c. and 39 u.c. devices**

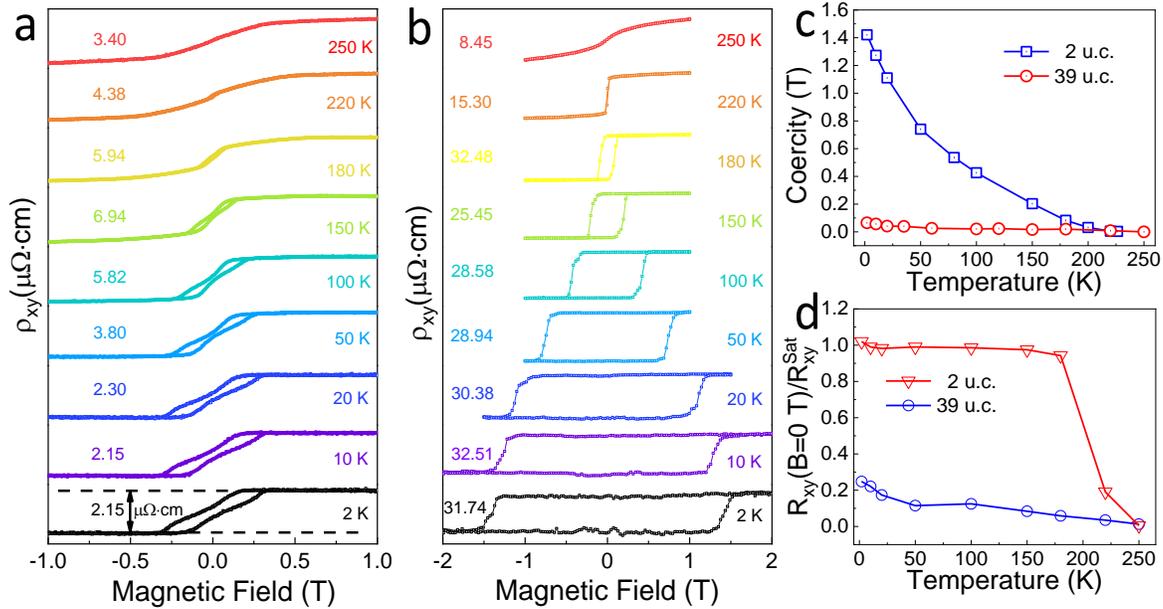

**Fig. S2 Detailed hysteresis behaviors of 2 u.c. and 39 u.c. F5GT devices. (a) (b)** Field dependent anomalous Hall resistivity for 2 u.c. and 39 u.c. F5GT devices at various temperatures. **(c)** Temperature dependence of coervities for 2 u.c. and 39 u.c. F5GT devices. For both devices the coercivity increases with the decreasing temperatures, while the coercity of 2 u.c. nanosheet is much larger than 39 u.c. one under 150 K. **(d)** Temperature dependent $R_{xy}(B=0T)/ R_{xy}^{SAT}$ values of 2 u.c. and 39 u.c. devices. Under 175 K the values stay high at around 1 for 2 u.c. device. The 39 u.c. device's values keeps lower than 0.1 over 50K and increases to over 0.2 at 2 K.

**Section S3. Definition of temperature dependent remanence curve**

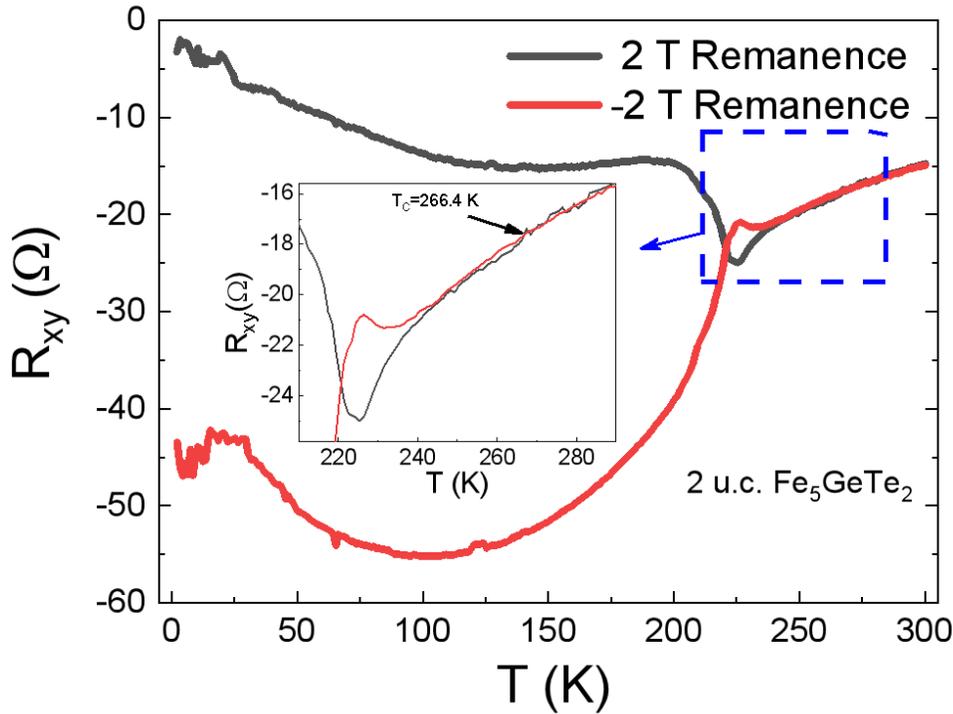

**Fig. S3 Definition of the temperature dependent remanence curve for 2 u.c. device.** During the measurement of 2 u.c. F5GT device, the sample was firstly cooled down to 2 K under a magnetic field of 2 T (-2T). Then the field slowly (5 Oe/s) scanned down to 0 Oe at 2 K. Finally we increase temperature to 300 K at 3 K/min and get the $R_{xy}$ vs T curve with 2 T (-2 T) remanence shown in the figure. The Remanence vs T curve in Fig.2B can be obtained from this figure by Remanence= $[R_{xy}(2T)-R_{xy}(-2T)]/2$. By telling the junction of remanence vs T curves of 2 T and -2 T where the remanence $R_{xy}(T)$ goes to zero as shown in the inset, we can get the value of $T_C$. According to this definition, the illustrated remanence curves of 2 T and -2 T saturation at 2 K meet at ~266.4 K, which is regarded as Curie temperature.

**Section S4. Protonic gating of other devices**

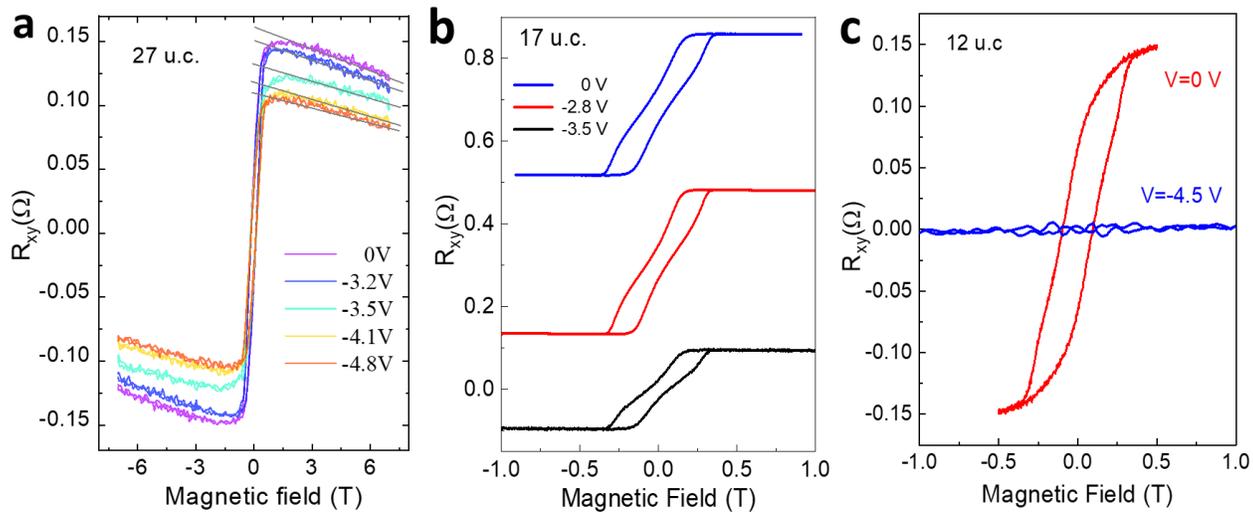

**Fig. S4. Protonic gating modulation of other 3 devices at 2 K.** (**a**) Magnetic field dependent $\rho_{xy}$ curves in saturation magnetization regime. The grey lines are the fitting curves to approximate the slop of each cure and derive the carrier density. (**b**) Anomalous Hall loops at gating voltages of 0 V, -2.8 V and -3.5 V for 17L F5GT nanosheet. The magnetization changes with the different gating voltages. (**c**) $R_{xy}(B)$ curves of 12 L devices. When $V_g$=4.5 V, the ferromagnetism is "turned off".

## Section S5. Definition of $T_k$

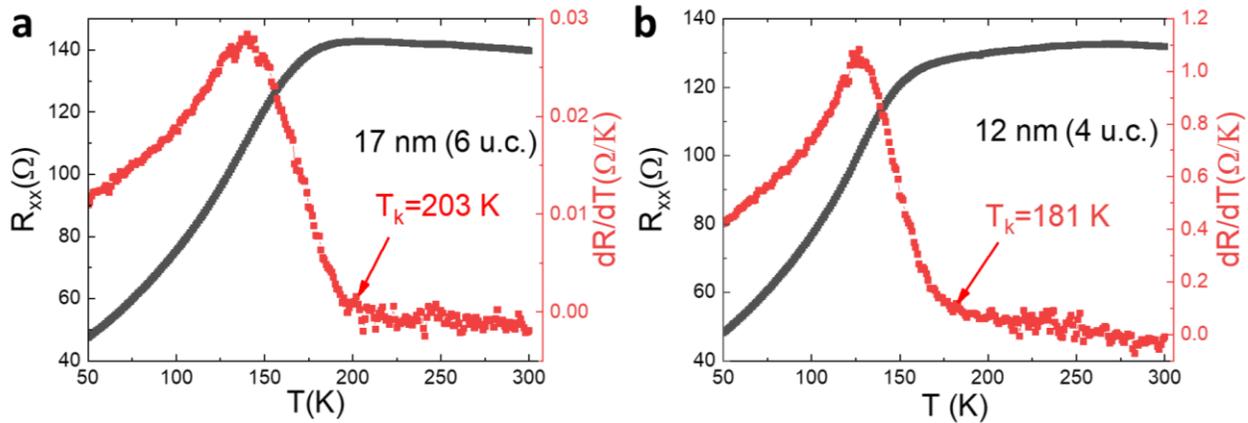

**Fig. S5 Definition of the Fe1 site magnetic transition temperature $T_k$.** Only 4 u.c and 6 u.c. devices are shown here as examples. For temperature dependent differentiate $R_{xx}$, the value firstly stay stable at higher temperatures, then the curves increase sharply. The turning point is regarded as $T_k$, where the magnetic transition of Fe1 site start. After reaching the peak point, the dR/dT values decrease, indicating the gradually evolution of $R_{xx}(T)$ curve as a standard metal behavior with rare spin disorder.

**Section S6. Angle dependent anomalous Hall measurement.**

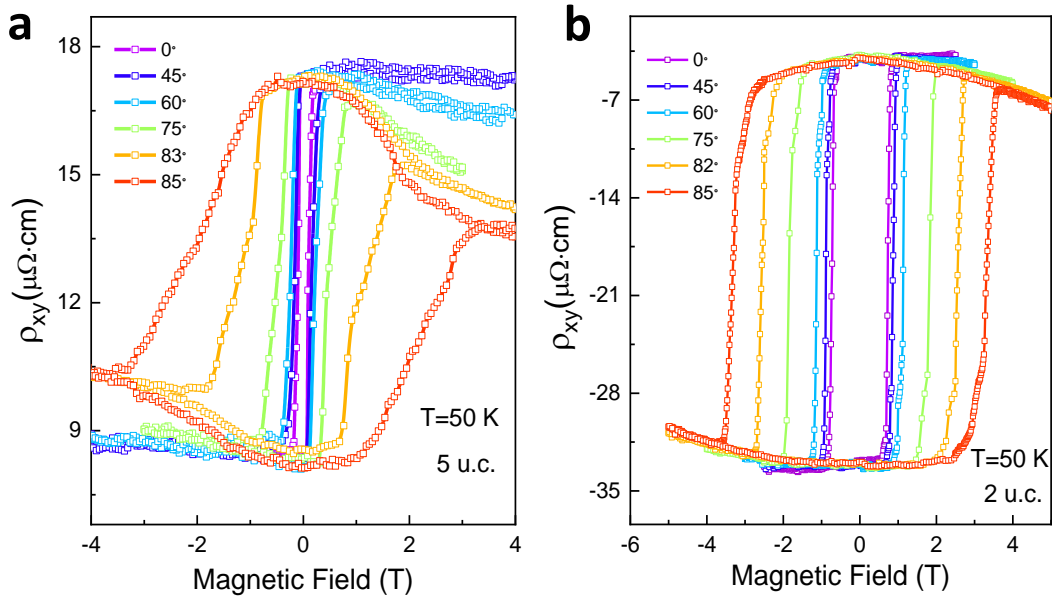

**Fig. S6. Angle dependent anomalous Hall measurement of 2 u.c. and 5 u.c. devices.** The tilt angle indicated is defined as the angle between the applied magnetic field and the direction perpendicular to the sample plane (magnetic easy axis). **(a)** $R_{xy}(B)$ curves at different angles between the applied magnetic field and the direction perpendicular to the surface of the nanoflake which is of 5 u.c.. At 0°, the surface of the nanoflake is perpendicular to the magnetic field. **(b)** $R_{xy}(B)$ loops at various tilt angles of the 2 u.c. F5GT device. The loops show a single domain behavior with strong perpendicular anisotropy.

## Section S7. Density functional theory calculation with different methods

|               | $a$ (Å) | $c$ (Å) | $M$ ($\mu_B$ / Fe atom) | (meV/m.f.) |
|---------------|---------|---------|-------------------------|------------|
| PBE           | 3.97    | 28.89   | 1.48                    | 1.45       |
| PBE+U         | 4.26    | 29.03   | 3.13                    | -63.38     |
| LDA           | 3.88    | 28.66   | 1.20                    | 0.17       |
| LDA+U         | 4.07    | 28.68   | 2.84                    | -1.42      |
| pw91          | 3.97    | 28.89   | 1.44                    | 0.62       |
| pw91+U        | 4.31    | 30.39   | 3.13                    | -90.23     |
| Experiment[1] | 4.04    | 29.19   | 1.80                    |            |

$$\Delta E = E_{FM} - E_{AFM}$$

**Table S1. Density functional theory calculations with different lattice parameters and methods.** Calculated lattice parameters ($a$ and $c$, in Å), the average magnetic moment per Fe atom ($M$, in the Bohr magneton $\mu_B$), and energy difference of bulk F5GT between FM and AFM coupling, with various exchange-correlation functionals. The vdW interaction is taken into account via Grimme's DFT-D3 with Becke-Jonson damping scheme, and effective Hubbard U approach is tested with the parameter U equals 4 eV. For comparison purpose, the available experimental data are listed as well.

**Section S8 HSE06 Band Structures and Hall conductivity evolution with electron doping**

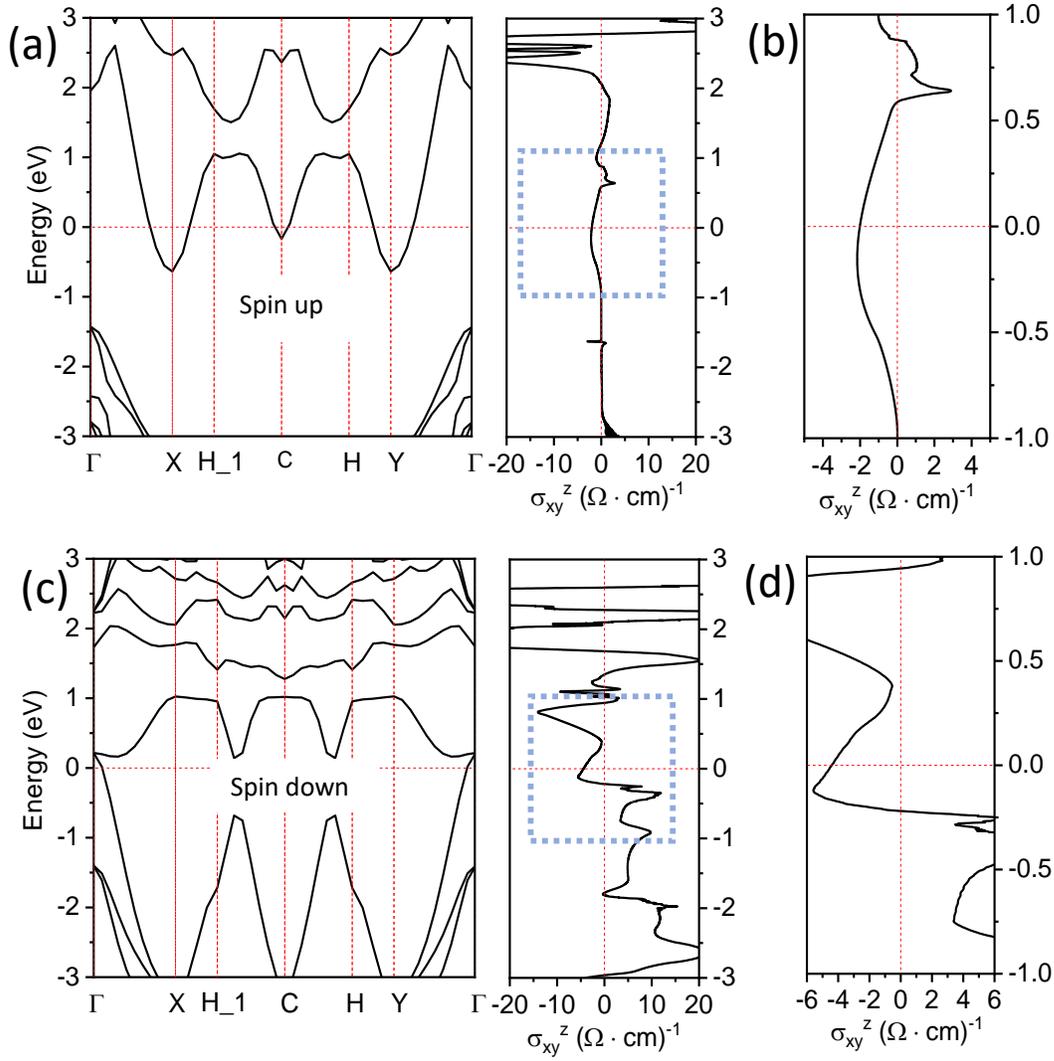

**Fig. S7 HSE06 band structures and Hall conductivity with electron doping.** Here, spin-down component with a relatively larger $\sigma_{xy}^z$ has a major contribution to the system compared to spin-up component. **(a)** Spin-up component of HSE06 band structure with corresponding Hall conductivity in F5GT. **(b)** Zoom-in of the blue dash square area in **(a)**. **(c)** Spin-down component of HSE06 band structure with corresponding Hall conductivity in F5GT. **(d)** Zoom-in of the blue dash square area in **(c)**.